\documentclass[preprint,aps]{revtex4}

\usepackage{graphicx}
\usepackage{dcolumn}
\usepackage{bm}

\begin{document}


\title{ Indirect Optical Absorption of Single Crystalline $\beta$-FeSi$_2$}

\author{Haruhiko Udono}
\email{udono@ee.ibaraki.ac.jp}
\author{Isao Kikuma}%
 
\affiliation{%
Department of Engineering, IBARAKI University, 4-12-1 Nakanarusawa, Hitachi, Ibaraki 316-8511, Japan
}%

\author{Tsuyoshi Okuno}
\author{Yasuaki Masumoto}%
\affiliation{
Institute of Physics, University of Tsukuba, 1-1-1 Tennohdai, Tsukuba, Ibaraki 305-8573, Japan
}%

\author{Hiroyuki Tajima}
\affiliation{
Institute for Solid State Physics, University of Tokyo, 5-1-5 Kashiwanoha, Kashiwa, Chiba 277-8581, Japan
}%


\begin{abstract}
We investigated optical absorption spectra near the fundamental absorption edge of $\beta$-FeSi$_2$ single crystals by transmission measurements. The phonon structure corresponding to the emission and absorption component was clearly observed in the low-temperature absorption spectra. Assuming exciton state in the indirect allowed transition, we determined a phonon energy of 0.031 $\pm$ 0.004 eV. A value of 0.814 eV was obtained for the exciton transition energy at 4K.
\end{abstract}

\maketitle

$\beta$-FeSi$_2$ is increasingly attracting attention as a suitable material for use in silicon based optoelectronic devices, due to its band gap being near the absorption minimum of quartz optical fibers~\cite{rf1, rf2}. 
Recently Leong {\it et al.} and Suemasu {\it et al.} fabricated light-emitting diodes operating at the wavelength of 1.5 -- 1.6 $\mu$m by introducing $\beta$-FeSi$_2$ particles into a silicon bipolar junction~\cite{rf3,rf4}.
Chu {\it et al.} also demonstrated the 1.57 $\mu$m electroluminescence (EL) at room temperature from sputter deposited $\beta$-FeSi$_2$ films on Si~\cite{rfChu}. However, the luminescence mechanism in $\beta$-FeSi$_2$ is not clearly understood, because the electronic structure of $\beta$-FeSi$_2$ is not clarified yet. The band gap nature, i.e., a direct or indirect gap, is also still controversial.

A number of experimental studies on the band gap nature of $\beta$-FeSi$_2$ have been performed by optical absorption measurements.
From the analysis of the energy dependence of the absorption coefficient, in most reports it is argued that $\beta$-FeSi$_2$ has a direct band gap~\cite{rf1, rf2, rf5, rf6, rf7, rf8, rf9, rf10, rf11}, but a few papers report an indirect gap lower than the direct one by some tens of meV.~\cite{rf12, rf13, rfTak}. 
The reported values of the band gap are 0.80 - 0.95 eV for direct gap and 0.7 - 0.78 eV for indirect one. The wide variation of the reported values suggests some uncertain factors existed in measured samples.

In order to study the band gap nature, optical transmission measurement using thick single crystalline samples is preferable because the absorption coefficient of crystals with an indirect energy gap is usually low.
However, no paper reports the optical transmission measurements of bulk $\beta$-FeSi$_2$ single crystals because of the difficulty of crystal growth. Recently, we have succeeded in growing large-sized $\beta$-FeSi$_2$ single crystals. In this paper, we report optical transmission measurements of $\beta$-FeSi$_2$ single crystals.


Single crystalline $\beta$-FeSi$_2$ ingots were grown by the temperature gradient solution growth (TGSG) method using Ga solvent. Details of the growth condition were described elsewhere~\cite{rfUd1, rfUd2, rfUd3}.
The crystals showed p-type conduction with a typical hole concentration of 1.5 $\times$ 10$^{19}$ cm$^{-3}$ at 300K and less than 1 $\times$ 10$^{16}$ cm$^{-3}$ at 25K. 
Crystals cut from grown ingots were ground using carborundum and polished using colloidal alumina. After the polishing, the surface of the crystals showed mirror-like face.
Optical transmission spectra were measured between 3.5 and 300K using a double-beam spectrophotometer (Hitachi U-4000). Reflection measurements were made at 300K using a UV-VIS-NIR microspectrophotometer (Nippon Bunko).
The absorption coefficient $\alpha$ was obtained by solving the following equation, assumed that the temperature dependence of reflectivity $R$ was negligible throughout the measured spectra region (0.7 --1.0 eV): 
\begin{eqnarray}
T = \frac{ I_T}{ I_T'} = \frac{ (1-R)^{2} \exp(-\alpha d)}{ 1-R^{2} \exp(-2\alpha d)}
\label{eq:one},
\end{eqnarray}
where $d$ is the thickness, $I_T$ the transmitted intensity and $I_T'$ the apparent transmitted intensity~\cite{rfGaP}.


  Figure 1 contains experimental data on the relationship between $\alpha$$^{1/2}$ and $h\nu$ obtained from transmission measurements on a crystal of thickness 0.0044 cm, recorded at six temperatures between 4 and 300K. The stepped structure characteristic of the intrinsic absorption edge of crystals with an indirect energy gap was observed in the low-temperature spectra. 
The energies $E_{P1}^{e}$ and $E_{P1}^{a}$ refer to the thresholds for structural components as defined in the 70K spectrum. The superscripts denote whether the phonon is emitted (e) or absorbed (a) during the optical absorption process. The strength of the phonon absorbed component with threshold at $E_{P1}^{a}$ decreased with decreasing the temperature, and the component was not present below about 40K. 
The difference between $E_{P1}^{e}$ and $E_{P1}^{a}$ was the same for each temperature within experimental uncertainty. Thus, $E_{P1}^{e}$ and $E_{P1}^{a}$ would be related to the thresholds of indirect transitions with a phonon emission or absorption.
From the observation of several absorption spectra in different samples, we found only one dominant phonon structure. Therefore, we analyze the spectra using one dominant phonon of energy $E_{ph}$.

We will assume that the Coulomb interaction between the excited electrons and holes is strong enough for the creation of free excitons to play a significant role in the optical absorption spectrum at the low temperature, as is so for Ge, Si and GaP~\cite{rfGaP, rfGe, rfSi}. Then, the optical absorption of indirect allowed transitions should be of the form~\cite{rf17}
\begin{equation}
\alpha(h\nu) = \frac{A}{\exp(E_{ph}/kT)-1}(h\nu - E_{gx} + E_{ph})^{1/2} + \frac{B\exp(E_{ph}/kT)}{\exp(E_{ph}/kT)-1}(h\nu - E_{gx} - E_{ph})^{1/2}
\label{eq:two},
\end{equation}
for the pair of component associated with a given phonon of energy$ E_{ph}$, which has the momentum required to take the electron from the valence-band maxima to the conduction-band minima. The energy $ E_{gx}$ is just the band gap energy minus an exciton binding energy. The quantities A and B are parameters containing the density-of-state effective masses of electrons and holes. $k$ is Boltzmann's constant.
According to Eq. (2), the strength of the absorption component $\alpha^{a}$ is proportional to the available phonon, and the strength depended on the temperature is given by 
\begin{equation}
\alpha^{a}(T) \propto \frac{A}{\exp(E_{ph}/kT)-1}
\label{eq:three}.
\end{equation}
Figure 2 shows the temperature dependence of absorption coefficient at the energy thresholds $E_{P1}^{e}$ for each spectra. The absorption coefficient at the energy thresholds increased with increasing the temperature as followed to Eq. (3). Thus, we obtained a phonon energy $E_{ph}$ = 0.031 $\pm$ 0.004 eV from the fitting curve using Eq. (3). Excellent agreement between the experimental absorption coefficient and the theoretical fitting provides convincing evidence that the absorption band comes from the phonon-assisted transition to the exciton state. Tassis {\it et al.} and Lange {\it et al.} reported the phonon band at 268.2 and 261 cm$^{-1}$, respectively, by infrared (IR) absorption measurements on $\beta$-FeSi$_2$ films on Si~\cite{rf6, rf12}. Guzzeti {\it et al.} pointed out that the highest-intensity peak was at about 250 cm$^{-1}$ for the Raman spectra measured on single crystalline $\beta$-FeSi$_2$~\cite{rf18}. Our phonon energy agreed with those reported values.
Different weak phonon peaks were also reported in the IR and Raman spectra. However, the phonon energy dominantly determining absorption profiles is believed to be 0.031eV.

We compared the experimental absorption spectra as the basic shape to the theoretical temperature dependence, on the assumption that the phonon energy of 0.031 eV is dominant during the optical transition in our crystals.
The results of such a comparison for 4K and 70K were shown in Fig. 3. So by considering only one phonon energy, rather good agreement between the experimental spectra and the calculated spectra is obtained. 
In our experiment, the ratio $B/A$ of the best-fitted spectra is not just unity but around 3.3.
From the fitting of the spectra, we obtained $E_{gx}$ = 0.814 eV at 4K and 0.810 eV at 70K. These values are approximately 0.1 eV lower than the values of the reported direct energy gap that were measured from $\beta$-FeSi$_2$ films on Si~\cite{rf6, rf7, rf9, rf10}. Based on the phonon-assisted transition probabilities, the small energy difference $\delta$E = 0.1eV and the phonon energy $E_{ph}$ = 0.031eV can gave the ratio $B/A$ = ($\delta$E + $E_{ph}$)$^{2}/ $($\delta$E - $E_{ph}$)$^{2}$ = 3.6 which is close to the experimental $B/A$. This agreement gives us the following definitive conclusion, $\beta$-FeSi$_2$ is an indirect band gap semiconductor, although the direct gap is very close to the indirect one.


  In conclusion, we have measured the optical absorption spectra near the fundamental absorption edge of $\beta$-FeSi$_2$ single crystals by transmission measurements. The stepped structure corresponding to the phonon emission and absorption is observed in the low-temperature absorption spectra below about 150K. We determined the exciton transition energy $E_{gx}$ is about 0.814 eV at 4K and about 0.810 eV at 70K, and also obtained a phonon energy $E_{ph}$ = 0.031 $\pm$ 0.004 eV from the analysis of the spectra. Our experimental results reveal that $\beta$-FeSi$_2$ has an indirect band gap.

  The authors would like to thank Professor Y. Maeda for fruitful discussions for Raman studies. This work was partially supported by the Murata Science foundation, the Grant-in-Aid for Encouragement of Yang Scientists (14750224) from the Japan Society for the Promotion of Science, and the Industrial Technology Research Grant Program in '02 (No. 02A22003a) from New Energy and Industrial Technology Development Organization (NEDO) of Japan.

\newpage 
\textbf{References}
\\

\newpage 
\begin{figure}[ht]
\vspace{-3 cm}
\hspace{-2 cm}
\includegraphics{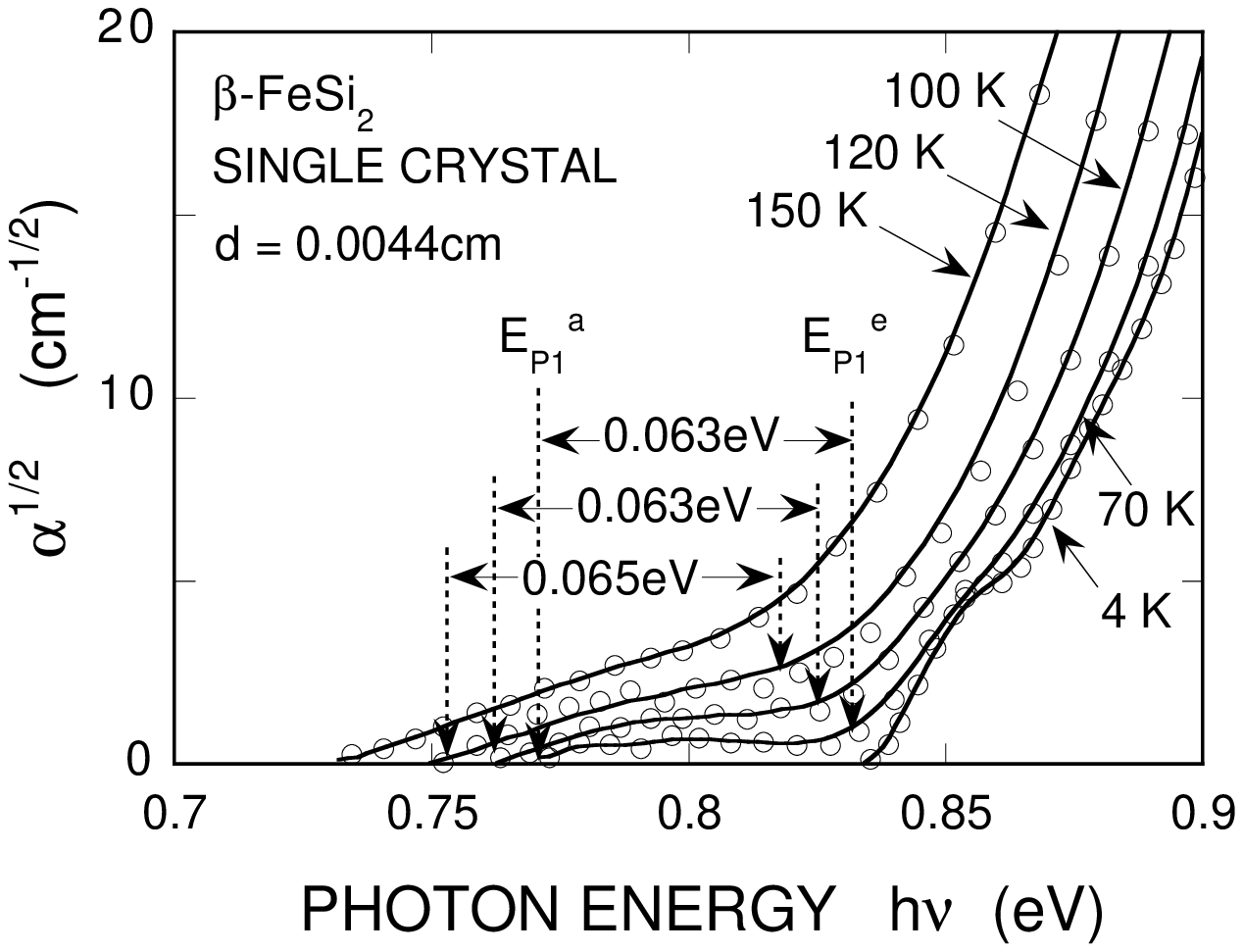}
\caption{ Fig. 1.   The low-level absorption spectra near the absorption edge of single crystalline $\beta$-FeSi$_2$ at various temperature. }
\end{figure}

\begin{figure}[ht]
\begin{center}
\includegraphics{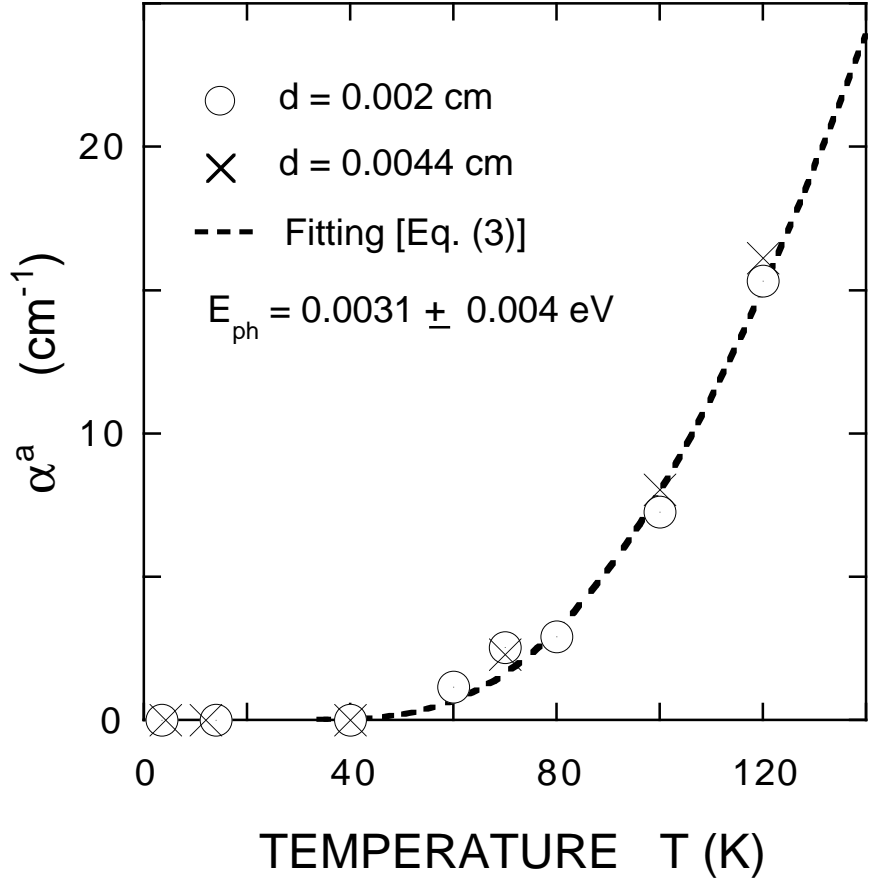}
\caption{ Fig. 2.  Temperature dependence of the strength of absorption component $\alpha^{a}$. The phonon energy $E_{ph}$ = 0.031 $\pm$ 0.004 eV was obtained from the fitting to Eq. (3).}
\end{center}
\end{figure}

\begin{figure}[ht]
\begin{center}
\includegraphics{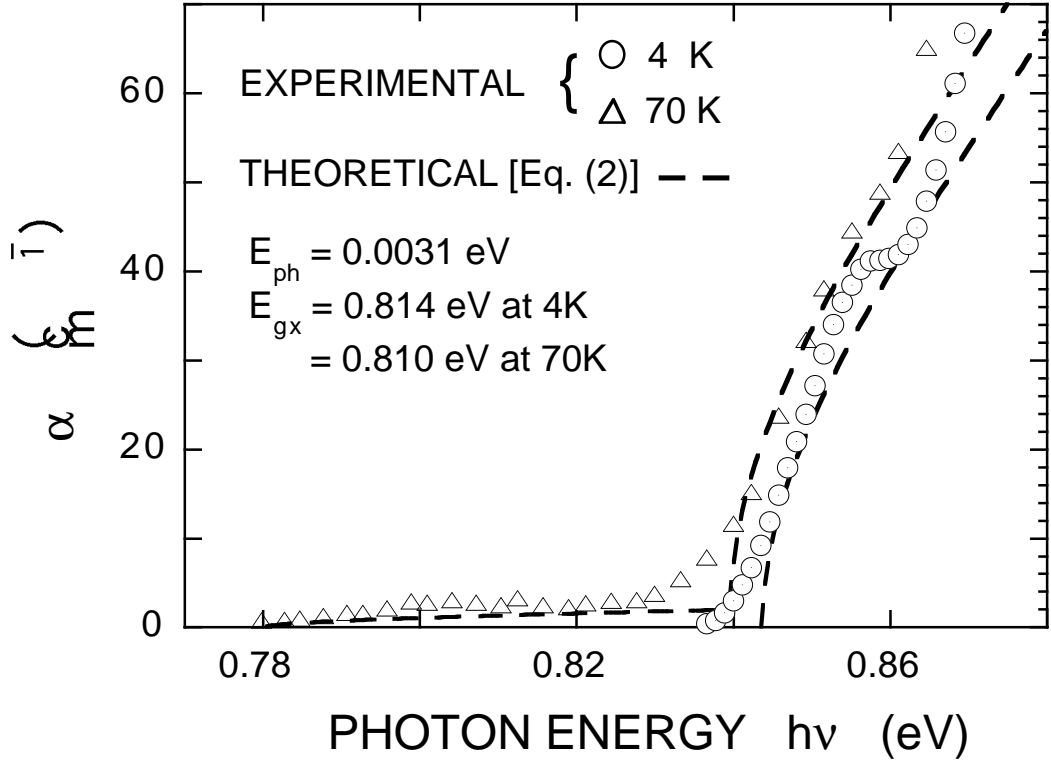}
\caption{ Fig. 3.  The edge-absorption spectra of $\beta$-FeSi$_2$ at 4 and 70K. The components associated with the phonon emission and absorption compared with the theoretical expression of Eq. (2), with the assumption of the one phonon energy of 0.031 eV.}
\end{center}
\end{figure}

\end{document}